\documentstyle[preprint,aps]{revtex} 
\begin{document}
\draft
\preprint{\today}
\title{Competing structural instabilities in cubic perovskites}

\author{W.~Zhong and David Vanderbilt}

\address{Department of Physics and Astronomy,
  Rutgers University, Piscataway, NJ 08855-0849}

\date{\today}
\maketitle

\begin{abstract}
We study the antiferrodistortive instability and its interaction
with ferroelectricity in cubic perovskite compounds.  Our
first-principles calculations show that coexistence of both
instabilities is very common.  We develop a first-principles scheme
to study the thermodynamics of these compounds when both
instabilities are present, and apply it to SrTiO$_3$.  We find that
increased pressure enhances the antiferrodistortive instability
while suppressing the ferroelectric one.  Moreover, the presence of
one instability tends to suppress the other.  A very rich $P$--$T$
phase diagram results.
\vskip 0.2truein\noindent
{\sl Submitted to Phys.\ Rev.\ Lett.}
\vskip 0.5truein
\end{abstract}
\narrowtext

The great fascination of the cubic perovskite structure is that it
can readily display a variety of structural phase transitions,
ranging from non-polar antiferrodistortive (AFD) to ferroelectric
(FE) and antiferroelectric in nature \cite{lines}.  The competition
between these different instabilities evidently plays itself out in
a variety of ways, depending on the chemical species involved,
leading to the unusual variety and richness of the observed
structural phase diagrams.  Moreover, all the phase transitions
involve only small distortions from the ideal cubic structure, and
are therefore appealing objects for experimental and theoretical
study.  However, our microscopic understanding of the chemical
origins of these instabilities and of their interactions is still
very limited.

Thus, there is a pressing need for accurate, chemically specific
investigations of the structural energetics of these compounds,
leading to a detailed understanding of the phase transition
behavior.  Previous phenomenological model Hamiltonian
approaches\cite{dove,pytt,cowl,pytt2} have been limited by
oversimplification and ambiguities in interpretation of experiment,
while empirical \cite{bilz} and nonempirical pair-potential methods
\cite{boyer} have not offered high enough accuracy.
First-principles density-functional calculations have been shown to
provide accurate total-energy surfaces for perovskites as regards
FE distortions \cite{cohen,singh,king1}.  However, to our
knowledge, there have been no previous first-principles studies of
AFD distortions, and therefore no detailed microscopic theories of
the phase transformation behavior.

Here, we build upon previous work in which a fully first-principles
scheme was used to study the FE transitions in BaTiO$_3$, leading
to an accurate microscopic understanding of the phase transition
sequence \cite{zhong2}.  In the present work, we develop a similar
approach which is capable of treating simultaneously the FE and AFD
degrees of freedom, allowing for the first time a detailed {\it
ab-initio} study of the phase behavior for perovskites in which
both instabilities are present.  We present systematic calculations
of the susceptibility against R-point zone-boundary AFD modes for a
set of eight compounds, demonstrating that the AFD instability is
very common.  Then, we briefly describe our first-principles scheme
for studying finite-temperature properties, and apply it to
SrTiO$_3$.  We study the evolution of the phonon instabilities with
temperature, and calculate the $P$--$T$ phase diagram.  In so
doing, we compute the interactions between the AFD and FE
instabilities, and expose their implications for the thermodynamic
properties.

The high-symmetry ABO$_3$ perovskite structure is simple cubic with
O atoms at the face centers and metal atoms A and B at the cube
corner and body center, respectively.  The two most common
instabilities result from a softening of either a zone-center polar
phonon mode (FE), or a non-polar zone-boundary mode (AFD) involving
rigid rotations of oxygen octahedra. These modes are illustrated in
the left and right insets, respectively, in Fig.~\ref{u-p}.
BaTiO$_3$ is a classical example of the first type, while the
best-known example of the second kind is the $T=105K$ transition in
SrTiO$_3$\cite{pytt2}, which results from a softening of a
$\Gamma_{25}$ phonon at $R$ [$(111)\pi/a$].

The stability of perovskite compounds against $R$-point phonon
distortions can be expressed in terms of a stiffness $\kappa^R=
(1/2)\partial ^2 E/ \partial \phi^2 $, where $\phi$ is the rotation
angle of the oxygen octahedra. To obtain $\kappa^R$, we perform
frozen phonon calculations using density-functional theory within
the local-density approximation (LDA) and Vanderbilt ultra-soft
pseudopotentials \cite{vand1}.  In Table \ref{table1},  we list
values of $\kappa^R$ for a set of eight compounds, calculated at
the experimental lattice constants\cite{exp_lat} as listed in
Ref.~\cite{king1}.  Negative values indicate instability to
$R$-point phonon distortions.

Table \ref{table1} shows that the tendency towards AFD instability
is strongly correlated with trends in ionic radii.  Such trends in
an ABO$_3$ compound are conventionally described by a tolerance
factor $t = (r_A + r_O)/\sqrt{2} (r_B + r_O)$.  Values for $t$ are
given in Table \ref{table1} using the ionic radii of
Ref.\ \cite{shan}.  We find that $\kappa_R$ is almost monotonic
with $t$, i.e., a larger A or a smaller B atom tends to stabilize
the cubic structure.  This simple behavior contrasts with the case
of the ferroelectric instability, where covalent interactions play
an important role \cite{zhong1}.

Inspecting Table \ref{table1}, we see that the two compounds
BaTiO$_3$ and KNbO$_3$ are clearly stable with respect to AFD
distortions, consistent with experimental observations. (Both
materials undergo a similar series of FE transitions.)  On the
other extreme, we find that CaTiO$_3$, PbZrO$_3$, and NaNbO$_3$
have a strong AFD instability. All three compounds are also
predicted to have FE instabilities\cite{king1}, consistent with the
observation of complex phase diagrams and high transition
temperatures in all three cases.  Finally, our calculations for
SrTiO$_3$, PbTiO$_3$, and BaZrO$_3$ show a weak AFD instability.
PbTiO$_3$ is observed to go through a weak unidentified transition
at $T$=180K\cite{kob} which could be AFD related.  BaZrO$_3$ is
observed to remain cubic down to $T$=0; the weak instability
predicted by our calculation could be suppressed by quantum
zero-point fluctuations.  For SrTiO$_3$, we predict a weak AFD
instability consistent with a low $T_c$ of 105K observed for its
cubic-to-AFD transition.

The above calculations indicate that coexistence of FE and AFD
instabilities is very common in perovskites.  To study the
consequence of such a situation, we have chosen to study the case
of SrTiO$_3$ in depth.  Our first-principles scheme can be
explained briefly as follows.  The energy is Taylor-expanded in
low-energy distortions, with expansion parameters determined from
LDA calculations. The resulting Hamiltonian is studied using Monte
Carlo (MC) simulations.  The low-energy distortions we included are
those connected with zone-center FE-like modes, zone-boundary
AFD-like modes, and strain.  To do this we construct a FE ``local
mode'' such that a uniform arrangement of local mode amplitudes
${\bf f}_l$ reproduces the softest zone-center $\Gamma_{15}$ (FE)
modes ($l$ is a cell index). Similarly, we construct an AFD local
mode (a local rotation of an oxygen octahedron) so that a staggered
arrangement of amplitudes ${\bf a}_l$ reproduces the
$\Gamma_{25}(R)$ mode.  Finally, the local strains are represented
in terms of a displacement vector ${\bf u}_l$.

Thus, we have three vector degrees of freedom {\bf f}$_l$, {\bf
a}$_l$, and {\bf u}$_l$ per cell.  The energy terms retained in our
Taylor expansion of the potential energy are: (i) on-site self
energy, up to quartic anharmonic order for {\bf f}$_l$ and {\bf
a}$_l$, and up to harmonic order for {\bf u}$_l$ (elastic energy);
(ii) harmonic intersite interactions between {\bf f}$_l$ (including
long-range dipole-dipole interactions) and between {\bf a}$_l$
(short-range only); and (iii) on-site coupling energy to the lowest
order between {\bf a}$_l$ and {\bf u}$_l$, between {\bf f}$_l$ and
{\bf u}$_l$, and between {\bf f}$_l$ and {\bf a}$_l$.  The
determination of the expansion parameters involves LDA calculations
for supercells containing up to 20 atoms with low symmetry, using
ultra-soft pseudopotentials \cite{vand1}.  The details of the
Hamiltonian, the first-principles calculations, and the values of
the expansion parameters will be presented elsewhere
\cite{zhong3}.

To obtain the structural and thermodynamic properties, we perform
MC simulations on an  $L\times L \times L$ cubic lattice with
periodic boundary conditions\cite{MC}.  The identification of
different phases can be made by monitoring the FE order parameter
${\bf f} (\Gamma)$ (the Fourier transform of ${\bf f}_l$ at ${\bf
k}=\Gamma$), and similarly the AFD order parameters ${\bf a}(R)$
and ${\bf a}(M)$ [$M=(110)\pi/a$].  ${\bf a} (M)$ is found to
remain small for SrTiO$_3$, and will not be discussed further.

We first investigate the ground-state structure for SrTiO$_3$ as a
function of hydrostatic pressure.  We find it convenient to run the
MC calculations at $L=4$ at $T$=0.1K (finite-size and hysteresis
effects are not important at low $T$).  The calculated order
parameters ${\bf a} (R)$ and ${\bf f} (\Gamma )$ are shown in
Fig.~\ref{u-p}.  Zero pressure in the figure corresponds to the
LDA-calculated equilibrium lattice constant, which is about 1\% too
small.  Since both the FE and AFD instabilities are sensitive to
lattice constant, comparison with the experimental phase diagram is
best made with the zero of the pressure axis shifted by
$P_0$=$-$5.4 Gpa (see arrow in Fig.~\ref{u-p}), the value which
restores the experimental lattice constant.  From Fig.~\ref{u-p},
we see that pressure has opposite effects on ${\bf a} (R)$ and
${\bf f} (\Gamma )$, and that as a function of pressure, the ground
state of SrTiO$_3$ can have four phases.  The cubic phase, which is
stable at high temperature, is not present. At high pressure, only
one component of  ${\bf a} (R)$ is non-zero, indicating an AFD
tetragonal structure ($I4/mcm$).  As $P$ is lowered, the
corresponding ($z$) component of ${\bf f} (\Gamma )$ becomes
non-zero, and the structure transforms to tetragonal with FE and
AFD ($I4cm$).  A further decrease of pressure creates a
low-symmetry monoclinic structure ($Pb$), in which all components
of ${\bf f} (\Gamma )$ and ${\bf a} (R)$ are non-zero.
 Finally, below $-$8 Gpa  the structure becomes FE rhombohedral
($R3m$).  We see that the coexistence of zone-center and
zone-boundary instabilities creates many different phases and
complicated structures, even at $T$=0.

At finite temperature, the behavior becomes even more interesting.
We first show our MC simulation for $P=P_0$ ($-$5.4Gpa) and
$L=12$.  We start at high temperature and decrease $T$ in small
steps, allowing the system to reach equilibrium at each step.  The
hysteresis and finite-size effects on the transition temperatures
are found to be negligible.  In Fig.~\ref{u-T}, we show the order
parameters ${\bf f} (\Gamma )$ and ${\bf a} (R)$ as a function of
$T$.  (Since the order parameter vectors may rotate, what we
actually show are the averaged maximum, intermediate, and minimum
components of each vector.) Naturally, the system is found to adopt
the cubic structure at high temperature.  As $T$ is reduced, a
transition to an AFD tetragonal structure occurs at 130K, as
indicated by a strong increase of $a_z (R)$.  A second transition
occurs at T=70K to a FE tetragonal structure, below which $f_z
(\Gamma)>0$.  At very low temperature (10K), the system transforms
to the low-symmetry monoclinic structure.

Comparing with experiment, we see that our cubic-to-FE(T)
transition at 130K corresponds very well to the observed one at
105K \cite{Tc}.  Our observations of additional transitions to
AFD+FE phases at 70K and 10K are not, however, in direct accord
with experiment.  Instead, they agree with the observed softening
of the FE polar phonons, which would extrapolate to a FE transition
close to 40K\cite{muller} or 20K\cite{viana}.  It has been
speculated that the absence of a true FE phase at $T$=0 is a result
of quantum fluctuations of atomic positions, leading to crossover
into a ``quantum paraelectric phase'' at very low temperature
\cite{muller,viana,mart}.  Our inability to obtain agreement
between the classical MC theory and experiment at $T$=0 lends
additional support to this conclusion.

To construct a $P$--$T$ phase diagram, we have carried out a series
of similar cooling-down simulations at different pressures.  As
shown in Fig.~\ref{P-T}, there are at least seven different phases
present.  At strong negative pressure, SrTiO$_3$ behaves rather
like BaTiO$_3$, with a cubic $\rightarrow$ tetragonal $\rightarrow$
orthorhombic $\rightarrow$ rhombohedral sequence of transitions on
cooling.  Increasing the hydrostatic pressure tends to stabilize
the AFD state and destabilize the FE one. The pressure coefficient
$dT_c/dP = 28 $K/Gpa at $P_0$ agrees well with the experimental
value of 25K/Gpa \cite{okai}.  At very high pressure, the system
undergoes a single transition to a tetragonal AFD structure.  In
the intermediate regime, the presence of both kinds of instability
creates a variety of phases, including the complicated monoclinic
structure.  The ordering of the FE and AFD transition temperatures
reverses $\sim$1.5 Gpa below $P_0$ (hashed area in
Fig.~\ref{P-T}).  In this critical region the AFD and FE transition
temperature change dramatically, and the system may possess some
interesting characteristics (e.g., extreme dielectric properties).

The dramatic reversal of the AFD and FE transition temperatures in
the hashed region of the $P$--$T$ phase diagram suggests the
presence of a competition between the two instabilities.  Our
first-principles theory confirms this and provides microscopic
insight into the competition.  The FE and AFD instabilities affect
each other mainly through the on-site anharmonic coupling, and
through their mutual coupling to the elasticity.  In SrTiO$_3$, the
on-site coupling is found to lead the FE and AFD modes to suppress
one another, while the coupling through strain tends to stabilize
tetragonal phases relative to other phases.  Our calculations show
that the former effect dominates.

One way of quantifying the importance of this competition is to
compare with what would happen if the FE or AFD degrees of freedom
were artificially frozen out.  We find that at $P_0$, the AFD
transition temperature would be 25\% higher if all ${\bf f}_l$ were
frozen to zero; conversely, the FE C--T transition would be 20\%
higher if all ${\bf a}_l$=0.  At $T$=0, freezing ${\bf f}_l$=0
reduces the cubic-to-AFD transition pressure from $-8$ to $-11.8$
Gpa, while freezing ${\bf a}_l$=0 increases the cubic-to-FE
transition from $-1.5$ to $0.8$ Gpa.  Thus, we see clearly that the
FE and AFD instabilities compete with and tend to suppress one
another.  Because of this competition the T(A,F) phase at $P_0$ is
only slightly more stable than the T(A) phase, even at $T$=0; the
energy surface relative to the FE distortion takes the form of a
very long and shallow double well.  This may help explain the
observed suppression of the ferroelectric phase by quantum
fluctuations \cite{muller,viana,mart}.

Much of the interesting portion of our phase diagram appears to the
left of $P_0$, i.e, at negative (inaccessible) physical pressures.
It would be interesting, therefore, to consider compounds such as
CaTiO$_3$ or NaNbO$_3$ which are FE at $P_0$, and study AFD
instabilities at elevated $P$.  While the exact details of our
phase diagram for SrTiO$_3$ should not be expected to carry over to
other perovskites, we expect the general features to persist,
especially the tendency of the FE and AFD instabilities to suppress
each other and the presence of complicated phase diagrams with
numerous phases.

In conclusion, we have performed a fully first-principles study of
the finite-temperature properties of perovskite compounds with both
FE and AFD type instabilities.  We find that AFD instabilities are
almost as common as FE ones in cubic perovskite compounds.  For
SrTiO$_3$, our calculated $P$--$T$ phase diagram shows that the FE
and AFD instabilities have opposite trends with pressure.  The
anharmonic on-site coupling between order parameters causes the AFD
and FE instabilities to tend to suppress one another.

We thank R.D.~King-Smith and K.M.~Rabe for discussions.  This work
was supported by ONR grant N00014-91-J-1184.  Partial supercomputing
support was provided by NCSA grant DMR920003N.

\newpage
\begin{table}
\caption{Calculated stiffness $\kappa^R$ of R-point AFD phonon mode
(in Hartree), and tolerance factor $t$,
for several perovskite compounds.  \label{table1}}
\begin{tabular}{lcclcc}
     & $\kappa^R$ & $t$ &          & $\kappa^R$ & $t$ \\ \tableline
 BaTiO$_3$ &  0.295 & 1.07 & SrTiO$_3$ & -0.042 & 1.01 \\
 KNbO$_3$  &  0.242 & 1.06 & NaNbO$_3$ & -0.133 & 0.97 \\
 BaZrO$_3$ & -0.021 & 1.01 & PbZrO$_3$ & -0.324 & 0.97 \\
 PbTiO$_3$ & -0.037 & 1.03 & CaTiO$_3$ & -0.375 & 0.97
\end{tabular}
\end{table}

\begin{figure}
\caption{$T$=0 order parameters vs.\ pressure for SrTiO$_3$.  Solid
and dashed lines denote Cartesian components of ${\bf f}(\Gamma)$
and ${\bf a}(R)$ respectively.  Phases are labeled by lattice
symmetry (R=rhom\-bo\-he\-dral, M=monoclinic, T=tetra\-go\-nal,
O=orthorhombic) and by instabilities present
(A=anti\-ferro\-distortive, F=ferroelectric).  Dotted lines denote
phase boundaries.  Vertical arrow indicates theoretical
pressure $P_0$ at which the lattice constant matches the
experimental $P$=0 one.  Left inset:  sketch of
displacements leading to R(F) phase (Sr is omitted for clarity).
Right inset: same for T(A) phase.
\label{u-p}}
\end{figure}

\begin{figure}
\caption{Order parameters of SrTiO$_3$ vs.\ $T$ at $P_0 =-$5.4
Gpa.  Upper panel: averaged largest, middle, and smallest Cartesian
components of ${\bf a}(R)$.  Lower panel:  corresponding quantities
for ${\bf f}(\Gamma)$.  Phase labels are the same as in Fig.~1.
\label{u-T}} \end{figure}

\begin{figure}
\caption{$P$--$T$ phase diagram of SrTiO$_3$.
Hashing indicates the critical region where dramatic changes
occur. Vertical dash-dotted line indicates the pressure $P_0$
corresponding to experimental $P$=0. Phase labels are the
same as in Fig.~1.
\label{P-T}}
\end{figure}


\begin{references}
\bibitem{exp_lat} Since $\kappa^R$ is pressure sensitive, using
experimental lattice constants avoids the error due to $\sim$1\%
underestimate of the LDA lattice constants.
\end{references}
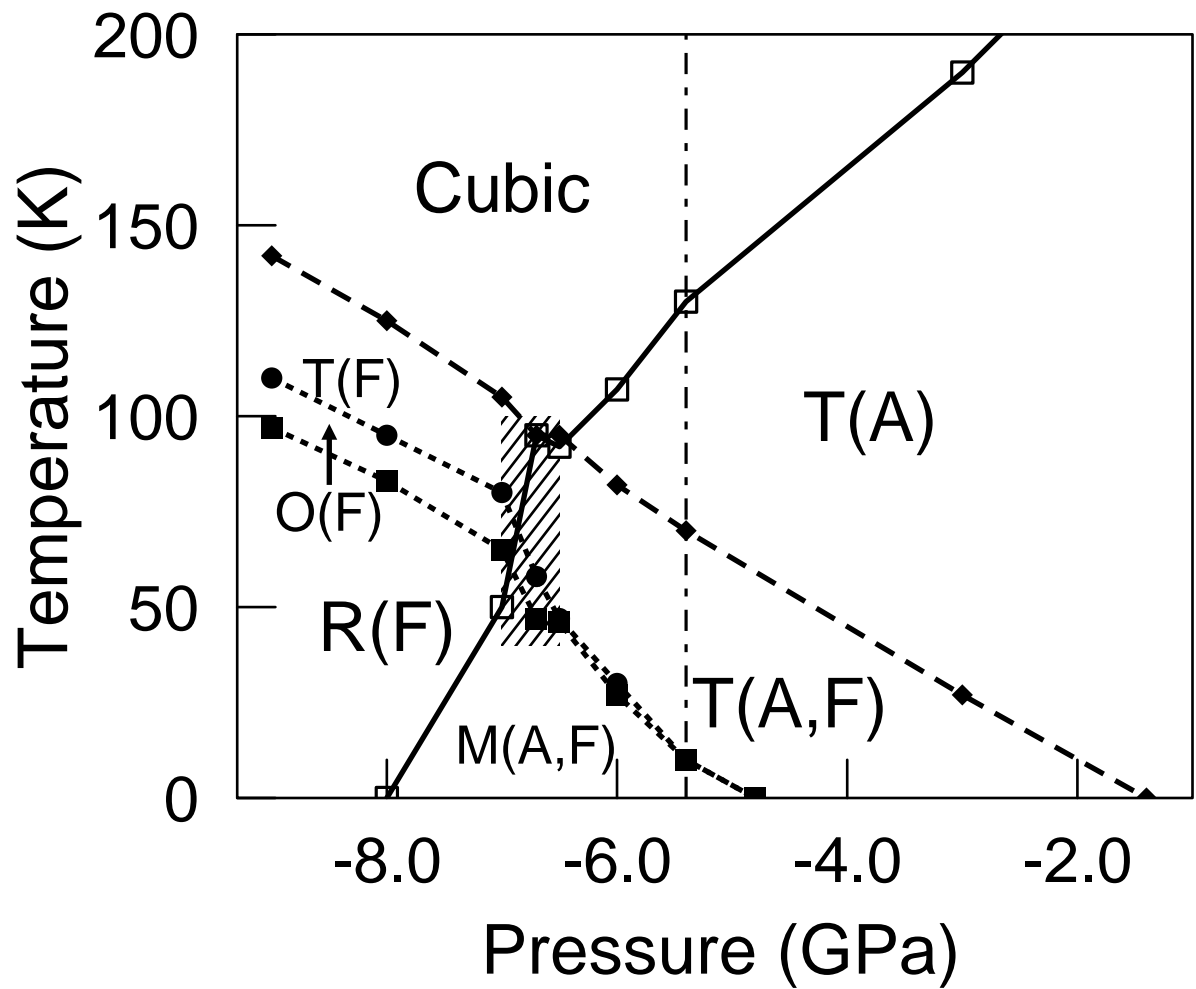
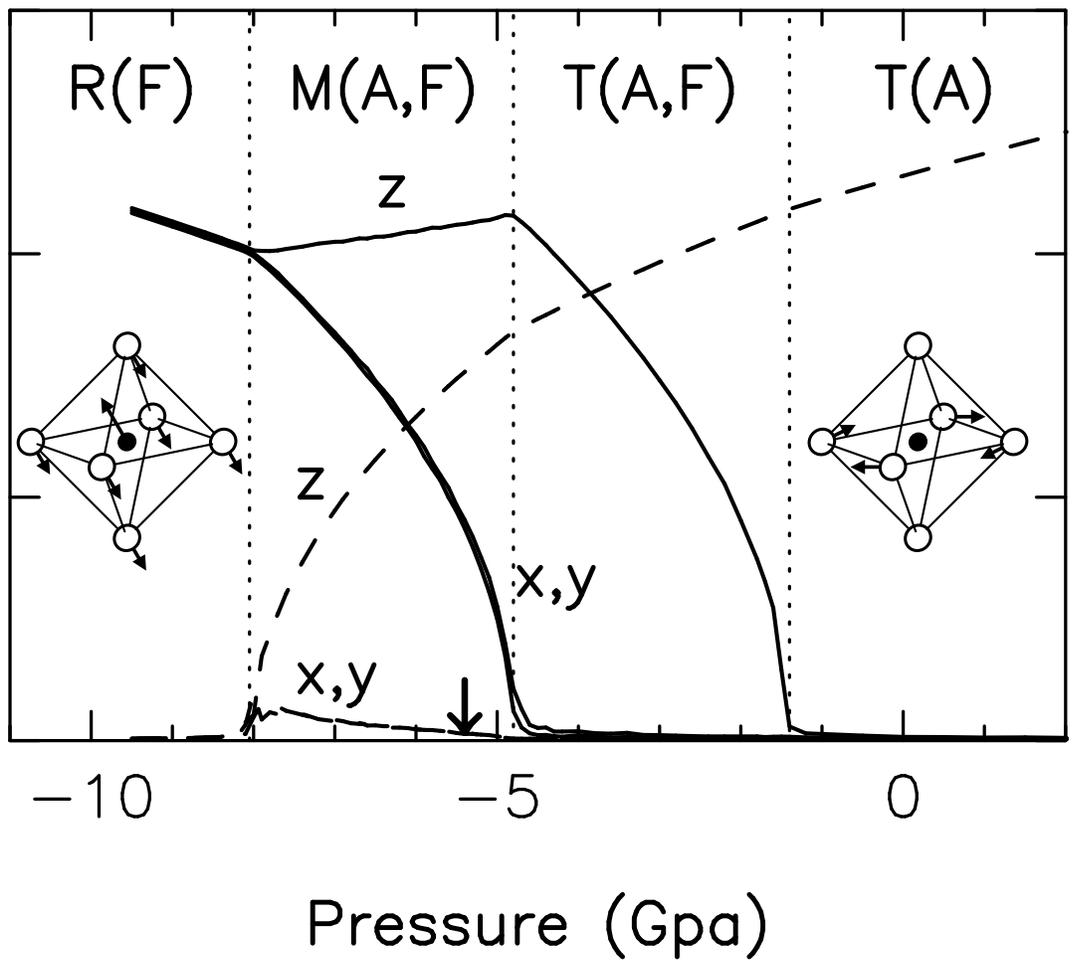
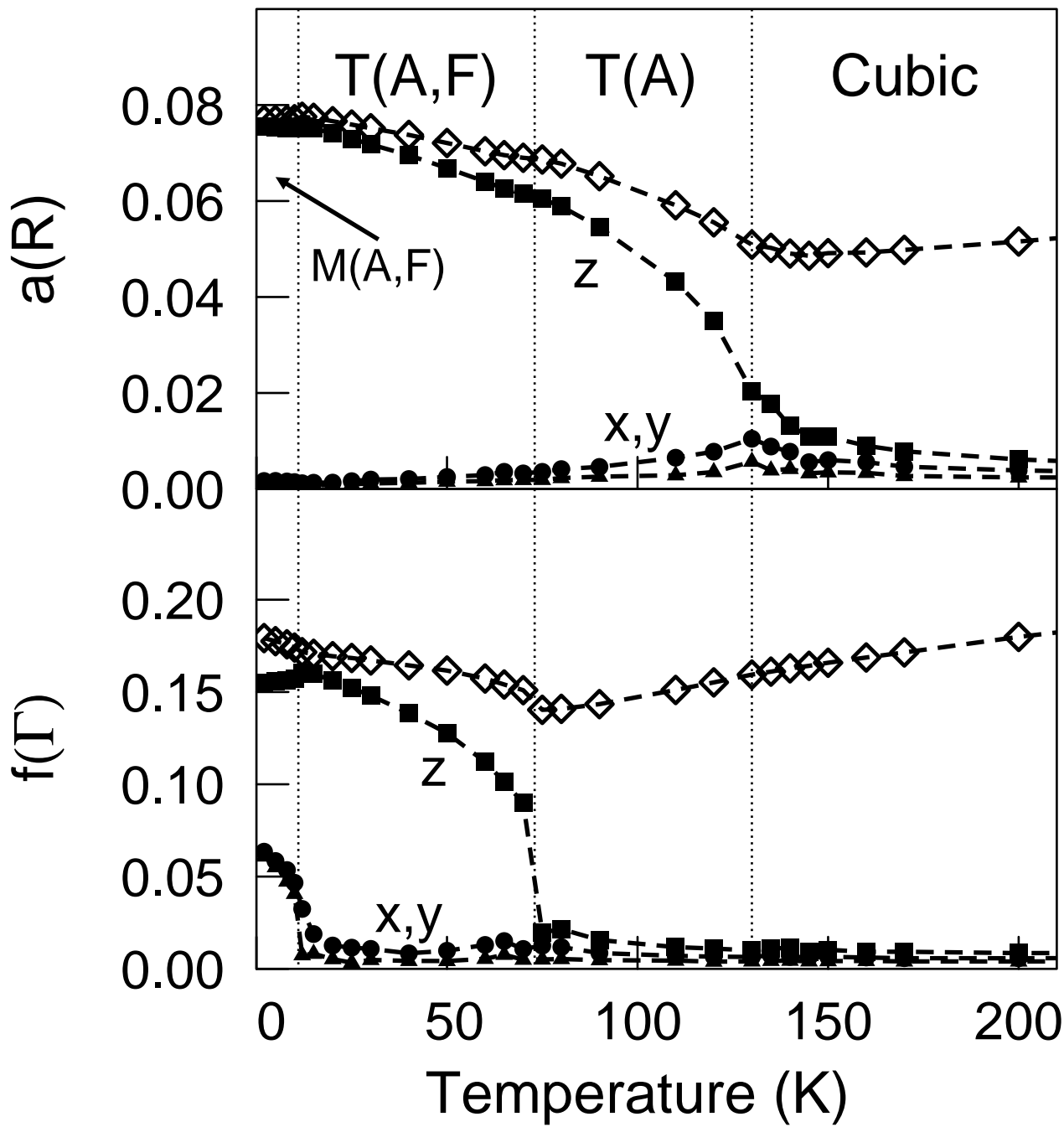
\end{document}